\documentstyle[preprint,aps]{revtex}
\begin{document}
\draft
\preprint{ }
\title{
Observation of Critical Phenomena and Self-Similarity \\ 
in the Gravitational Collapse of Radiation Fluid}
\bigskip
\author{Charles R. Evans and Jason S. Coleman}
\address{
Department of Physics and Astronomy \\
University of North Carolina, Chapel Hill, North Carolina 27599}
\date{\today}
\maketitle
\begin{abstract}

We observe critical phenomena in spherical collapse of radiation fluid.  
A sequence of spacetimes $\cal{S}[\eta]$ is numerically computed, 
containing models ($\eta\ll 1$) that adiabatically disperse and 
models ($\eta\gg 1$) that form a black hole.  Near the critical point 
($\eta_c$), evolutions develop a self-similar region within which collapse 
is balanced by a strong, inward-moving rarefaction wave that holds $m(r)/r$ 
constant as a function of a self-similar coordinate $\xi$.  The self-similar 
solution is known and we show near-critical evolutions asymptotically 
approaching it.  A critical exponent $\beta \simeq 0.36$ is found for 
supercritical ($\eta>\eta_c$) models.

\bigskip\bigskip
\noindent PACS numbers: 04.20.Dw, 04.25.Dm, 04.40.Nr, 04.70.Bw
  
\end{abstract}
\newpage

New solutions of Einstein's equations at the threshold of black hole 
formation have recently been discovered.  These solutions exhibit nonlinear 
dynamical behavior closely analogous to critical phenomena, and include 
power-law behavior, {\it discrete} scaling relations and a form of 
universality.  All of these features were discovered in 
Choptuik's~\cite{chop93} studies of spherical collapse of massless scalar 
field.  Abrahams and Evans~\cite{ae93} found strikingly similar behavior in 
axisymmetric collapse of gravitational waves, indicating the behavior is a 
generic feature of gravity.  Confirmation has recently been made~\cite{gpp93} 
of some of Choptuik's results.  In this {\it Letter,} we report observation 
of critical phenomena in spherical collapse of radiation fluid and show 
the asymptotic approach of near-critical spacetimes to a self-similar 
solution near the center of collapse.  This new result is fundamental in 
two ways; the self-similar solution has been separately and precisely 
calculated~\cite{evans93}, and it exhibits {\it local} self-similarity.

The process for finding the threshold of black hole formation and critical 
phenomena in gravitational collapse has been described 
elsewhere~\cite{chop93,ae93,ae93b,eard93}.  Adopting a physical model with 
metric {\bf g}, matter fields $\phi_A$, stress-energy {\bf T}$(\phi_A)$, 
and symmetries, the future development is computed of elements of a family 
of Cauchy data, distinguished by a dimensionless strength-parameter $\eta$ 
and a length (or mass) scale $r_0$.  Subcritical models ($\eta\ll 1$) have 
smooth future development, dispersing all mass-energy to infinity, while a 
black hole appears in supercritical models ($\eta\gg 1$) with only some 
mass-energy escaping.  As $\eta$ is tuned in a search for the onset (at 
$\eta_c$) of black hole formation, black hole mass (for $\eta>\eta_c$) 
diminishes as a power law: $M_{\rm BH} = K(\eta-\eta_c)^\beta$, with 
$\beta>0$.  In both scalar field collapse~\cite{chop93} and vacuum 
collapse~\cite{ae93,ae93b}, simulations give $\beta\simeq 0.37$, results 
that appear to be universal (i.e., independent of the parameterization of 
the initial data).  

By implication, tuning the parameter to just the right value ($\eta=\eta_c$) 
creates arbitrarily small black holes and regions of arbitrarily high 
spacetime curvature from smooth Cauchy data.  Hence, it is 
argued~\cite{eard93} that the threshold of black hole formation is also the 
threshold of naked singularity formation.  Spacetimes with $\eta=\eta_c$ are 
termed {\it precisely critical.}

In any physical model, many families of Cauchy data may exist with a 
critical point and a precisely-critical spacetime.  Each family of data 
will have a characteristic proper length scale $r_0$.  Related to this scale 
is the limiting interval, $\lim {|T|}=|T_0|$, as $\eta-\eta_c\to 0^{+}$,
of proper time $T$ of the observer $\cal{O}$ centered in the collapse.  Let 
$T=-T_0$ correspond to the initial data.  Further, let $r$ be a proper 
spatial distance from $\cal{O}$.  The central region of collapse $\cal{R}$ 
(in spacetime) exists for $r\ll r_0$, $|T|\ll |T_0|$.  Different 
precisely-critical spacetimes can be normalized, using scale invariance, to 
a common length scale $r_0$.  These spacetimes will still differ on scales 
$T\sim -T_0$ and $r\sim r_0$ but approach a (physical-model-specific but 
otherwise unique) self-similar solution within $\cal{R}$.  Within $\cal{R}$, 
all knowledge of the initial data is lost up to a single, dimensional 
constant.

Previous critical systems have displayed discrete self-similarities.  Let 
$q$ denote dimension and let length have $q=1$.  Allow the metric {\bf g} to 
carry dimension $q=2$~\cite{eard74}.  For $\eta=\eta_c$, within $\cal{R}$ a 
map will exist under which ${\bf g}\to{\bf g}^{\prime}=e^{-2\Delta}{\bf g}$, 
for a fixed value of $\Delta$.  These are echoes resulting purely from 
nonlinearities.  The scale factors $\Delta$ are pure numbers that emerge 
from the dynamics, with $e^{\Delta}\simeq 30$ in scalar wave 
collapse~\cite{chop93} and $e^{\Delta}\simeq 1.8$ in gravitational wave 
collapse~\cite{ae93,ae93b}.  

An analytic explanation of these phenomena would be important.  This has 
been sought for scalar field collapse but the discrete self-similarity 
proved to be an impediment.  The effort made obvious, though, that a local 
self-similarity would be more powerful, and suggested examining another 
system: perfect fluid collapse.

Imagine a spherical configuration of radiation fluid confined within areal 
radius $r_0$.  Let the total gravitational mass be $M$.  Define a 
dimensionless control parameter $\eta=2M/r_0$~\cite{units}.  When 
$\eta\ll 1$, pressure dominates and should cause adiabatic expansion and 
dispersal of the fluid.  When $\eta\agt 1$, gravity should overwhelm 
pressure and spur the formation of a black hole.  The onset of black hole 
formation should occur at some $\eta_c\sim 1$.  One might anticipate, with 
a fluid, that any self-similarity emerging near $\eta_c$ will be local.

A successful attempt was made to find a self-similar solution representing 
the threshold of black hole formation in perfect fluid 
collapse~\cite{evans93}.  Details of this solution will be given elsewhere; 
here we only describe the self-similar ansatz and a few numerical results.  
We assume a perfect fluid $T^{\mu\nu}=\rho U^{\mu}U^{\nu}+pg^{\mu\nu}$, with 
total energy density $\rho$, isotropic pressure $p$ and four-velocity 
$U^{\mu}$ satisfying $U_{\mu}U^{\mu}=-1$.  A relativistic equation 
of state $p=(\gamma-1)\rho$ is adopted.  In this paper, we specialize to 
radiation fluid $p={\frac{1}{3}}\rho$.  Adopting (dynamical) Schwarzschild 
coordinates, the line element is
\begin{equation}\label{lineel}
ds^2 = -\alpha^2 dt^2 + a^2 dr^2 + r^2 d\Omega^2 ,
\end{equation}
with $\alpha(r,t)$ the lapse function and $a(r,t)$ the radial metric 
function.  Constant proper sound speed $c_s=\sqrt{p/\rho}=\sqrt{\gamma-1}$ 
suggests the self-similar ansatz: $a(\xi)$, $U^r(\xi)$, 
$\rho(r,t)=\Omega(\xi)/4\pi r^2$, $\alpha(r,t)=nrN(\xi)/t$, with 
self-similar coordinate $\xi=r/C(-t)^n$.  Geometrically, 
${\bf w}=(1/n)t\partial/\partial t +r\partial/\partial r$ is the infinitesimal 
generator of the homothetic motion: $\pounds_{\bf w}{\bf g}=2{\bf g}$, 
$\pounds_{\bf w}{\bf T}=0$.  The similarity exponent $n$ is 
initially unknown, making the problem (in these coordinates) a type-2 
self-similarity~\cite{sedov}.  With this ansatz, the fluid and gravitational 
equations reduce to ordinary differential equations.  These are solved by 
demanding regularity at $\xi=0$ ($r=0$, $t<0$), $\xi=\infty$ ($r>0$, $t=0$), 
and along the limiting, ingoing acoustic characteristic (sonic point) at 
$\xi=1$, across which the fluid might otherwise be discontinuous.  The 
system is over-determined unless $n$ is taken as an eigenvalue of the 
nonlinear system.  An analogous type-2 similarity problem is Guderley's 
shock implosion problem~\cite{gud}.

The similarity exponent was found to be $n\simeq 1.1485$ ($\gamma=4/3$).  The 
solution is depicted in Fig.~1.  A physical singularity exists at $r=0$, 
$t=0$, since $R_{\mu\nu}R^{\mu\nu}={256\over 3}\pi^2\rho^2 \sim T^{-4}$ and
central proper time $T\sim (-t)^n$.  Notice that $a$ does not approach unity 
as $\xi\to\infty$, but rather approaches $a\simeq 1.07$.  Likewise, mass 
grows linearly with distance, with $m(r)/r\to 0.0596$ and 
$\Omega\to 9.56\times 10^{-3}$, so the solution is not asymptotically flat.  
An asymptotically-flat spacetime would require truncating the 
self-similarity at some large $r$~\cite{op90}.  Nonetheless, the 
self-similar solution is anticipated to represent the asymptotic behavior 
within $\cal{R}$ (i.e., as $r\to 0$, $t\to 0$). 

To confirm these expectations, we compute a series of models of 
spherically-symmetric collapse of radiation fluid, searching for a critical 
point and critical behavior.  We adopt polar time slicing and radial
gauge, retaining the form (\ref{lineel}) of the line element, and assume
a radiation fluid.  We define $W\equiv\alpha U^t$ and $U\equiv a U^r$, with 
which the velocity normalization becomes $W^2=1+U^2$.  Then, the equations 
of motion of the fluid, $\nabla_{\nu}T^{\mu\nu}=0$, are
\begin{eqnarray}
&&{\partial\over\partial t}\left(a\rho W\right) 
+{1\over r^2}{\partial\over\partial r}\left(r^2 \alpha \rho U\right) 
+p\left[
{\partial\over\partial t}\left(aW\right) 
+{1\over r^2}{\partial\over\partial r}\left(r^2 \alpha U\right)\right]=0 , 
\\
\nonumber\\
&&{\partial\over\partial t}\left[a^2(\rho+p)WU\right] 
+{1\over r^2}{\partial\over\partial r}
\left[r^2 \alpha\, a\, (\rho+p)U^2\right] 
\nonumber\\
\nonumber\\
&&
\qquad\qquad\qquad\qquad
+\alpha\, a\, (\rho+p)\left(
W^2{1\over\alpha}{\partial\alpha\over\partial r}
-U^2{1\over a}{\partial a\over\partial r}\right)
+\alpha\, a\, {\partial p\over\partial r} = 0 ,
\end{eqnarray}
and the pressure is $p={\frac{1}{3}}\rho$.  In these coordinates, the gravitational 
field equations ${G^r}_r=8\pi{T^r}_r$ and ${G^t}_t=8\pi{T^t}_t$ become
\begin{eqnarray}\label{hamcon}
&&{1\over a}{\partial a\over\partial r}+{1\over 2r}\left(a^2-1\right)
= 4\pi r a^2 \rho \left[\gamma W^2 - \gamma +1\right] ,
\\
\nonumber\\ \label{lapseeq}
&&{1\over\alpha}{\partial\alpha\over\partial r}-{1\over 2r}\left(a^2-1\right)
= 4\pi r a^2 \rho \left[\gamma U^2 + \gamma -1\right] ,
\end{eqnarray}
with $\gamma=4/3$ adopted in the models we discuss here.

To construct a parameterized family of evolutions, we choose Cauchy data 
by specifying that the fluid be instantaneously at rest $V\equiv U/W=0$ and 
have an energy density profile 
\begin{equation}
\rho = {1\over 2}\,\pi^{-3/2}\, r_0^{-2}\, \eta\, \exp(-r^2/r_0^2) ,
\end{equation}
parameterized by $\eta$ and $r_0$.  The total gravitational mass is
$M={\frac{1}{2}}\eta r_0$, so that $\eta=2M/r_0$ is a dimensionless 
measure of the 
strength of the initial gravitational field, while $r_0$ is the typical 
initial length scale.  Equations (\ref{hamcon}) and (\ref{lapseeq}) are 
solved to complete the Cauchy data.

The dynamical behavior of these models is consistent with expectations.  
For $\gamma=4/3$ and this family of Cauchy data, the critical point is 
found in a bisecting search to be near $\eta_c\simeq 1.0188$.  For 
$\eta>\eta_c$, the {\it impending} formation 
of a black hole is observed.  As is well-known, polar time slices avoid 
penetrating regions of trapped surfaces~\cite{polar} and asymptotically 
approach the apparent horizon in collapsing spacetimes.  When this occurs 
we have clear evidence of a black hole forming.  Fig.~2 illustrates the form 
of the mass function $m(r)={\frac{1}{2}} r(1-a^{-2})$ late in a supercritical 
evolution.  The ``kink'' where $r\simeq 2m(r)$ locates the surface of the 
black hole and determines the hole's mass $M_{\rm BH}$.  Immediately beyond 
the kink, $m(r)$ is nearly constant indicating an evacuated region that 
forms around the black hole.  At larger radii, additional mass is evident, 
representing the fluid which did not collapse and is expanding outward.  

Black-hole mass $M_{\rm BH}$ is found to be well described by a power law 
(see Fig.~3)
\begin{equation}\label{powerlaw}
M_{\rm BH} \simeq K (\eta-\eta_c)^{\beta} ,
\end{equation}
with a critical exponent $\beta\simeq 0.36$.  This critical exponent value 
is numerically indistinguishable from those found in scalar wave 
collapse~\cite{chop93} and gravitational wave collapse~\cite{ae93}, 
doubtlessly reflecting a deep property of the gravitational field equations.  

The fluid is initially everywhere out of equilibrium.  In near-critical 
models, the fluid in the inner region collapses while that in the outer 
region accelerates and expands outward.  The collapsing central region is
chased by a strong, ingoing rarefaction wave.  As the radius $R(t)$ of the 
inner region diminishes, the rarefaction wave causes the mass to diminish 
also, keeping $m(R(t))/R(t)$ nearly constant.  The radius $R(t)$ of transition
between collapse and expansion (edge of the rarefaction wave) decreases 
by several orders of magnitude in very-near-critical models.  On fine 
scales, $r\ll r_0$, the fluid and gravitational field assume a self-similar 
form (see Fig.~4).  Furthermore, the self-similar form being approached 
{\it is} the locally self-similar solution outlined above.  As Fig.~4 shows, 
ultimately the self-similarity of the collapse is broken if $\eta\neq\eta_c$.
Only a precisely-critical model would approach the self-similar solution
and retain this dependence as $r\to 0$, $t\to 0$.
Near- but not precisely-critical spacetimes develop a self-similar region 
on scales $r\ll r_0$ but then lose self-similarity on a scale determined by 
proximity in parameter space to the critical point: 
$r_1\simeq K|\eta-\eta_c|^{\beta}$.  Self-similarity will only be apparent 
if $r_1\ll r_0$.

Approach to self-similarity within a region spanning two disparate scales 
is a familiar concept in hydrodynamics termed ``intermediate 
asymptotics''~\cite{bz}.  The cardinal example is the modified Sedov-Taylor 
blast wave problem.  Besides the usual Sedov-Taylor parameters of $E_0$, the 
blast wave energy, and $\rho_0$, the ambient gas density, the modified 
problem has a length scale $r_0$, within which the energy $E_0$ is 
arbitrarily distributed initially, and has a small pressure $p_0$ in the 
ambient gas.  This small pressure introduces a second, new length scale: 
$r_1 = (E_0/p_0)^{1/3}\gg r_0$.  These two additional dimensional parameters 
break the self-similarity of the blast wave on length scales $r\alt r_0$ and 
$r\agt r_1$ but on intermediate scales the solution asymptotically 
approaches the Sedov-Taylor self-similar form.  There is a direct analogue
in the behavior of near-critical spacetimes on scales between $r_0$ and
$r_1$. 

The self-similar solution determines the asymptotic properties of 
precisely-critical evolutions.  Linear stability analysis should reveal the 
perturbative response of solutions nearby $\eta_c$ in parameter space and 
thereby perhaps provide an estimate of the critical exponent $\beta$.  The 
self-similar solution, and therefore any precisely-critical evolution, 
apparently contains a naked singularity.  However, the precisely-critical 
evolutions are likely to be a set of measure zero as there is extreme 
sensitivity to the initial conditions near $\eta_c$.  Furthermore, the 
dependence of the lapse along $\xi=\infty$ ($t=0$) is 
$\alpha\sim r^{1-1/n}\sim r^{0.129}$, indicating the singularity to be 
shrouded in infinite redshift.

C.R.E.~acknowledges helpful conversations with A.~Abrahams, M.~Choptuik, 
R.~Price and J.~York.  This research was supported by NSF grants PHY 
90-01645, PHY 90-57865 and ASC 93-18152, which includes support from ARPA.  
C.R.E.~thanks the Alfred P.~Sloan Foundation for research support and 
thanks the Aspen Center for Physics for hospitality during summer 1993 where 
key aspects of this work were initiated.  Computations were performed at 
the North Carolina Supercomputing Center.

\newpage

\begin{figure}
\caption{Self-similar radiation fluid collapse.  Fluid and metric variables 
are plotted versus self-similarity coordinate $\xi=r/C(-t)^n$ (with 
$n\simeq 1.1485$): (1) radial metric function $a(\xi)$ (solid), (2) velocity 
$V(\xi)=aU^r/\alpha U^t$ (dotted), (3) $\Omega(\xi)=4\pi r^2\rho$ (dashed), 
and (4) lapse function $\alpha=nrN(\xi)/t$ (dot-dashed).  Note the 
collapsing interior and expanding exterior regions.}  
\label{fig1}
\end{figure}

\begin{figure}
\caption{The mass function $m(r)$ (solid) at a late time in a supercritical 
evolution.  Polar time slices limit outside the horizon producing the kink 
near where the function approaches the curve (dotted) $r=2m(r)$, locating 
the surface of the black hole and determining the hole's mass.  At late 
times, the fluid bifurcates and forms an evacuated region immediately beyond 
the hole, indicated by the decrease of $\Omega=4\pi r^2\rho$ (dashed) 
outside the hole.  At larger radii, escaping fluid expands outward.}  
\label{fig2}
\end{figure}

\begin{figure}
\caption{Critical behavior of black hole mass.  For $\eta>\eta_c$, 
$M_{\rm BH}$ is determined and fit to a power law of the 
form (8).  Here log of $M_{\rm BH}$ is plotted versus log of the critical 
separation $\eta-\eta_c$.  The power-law fit is indicated by the solid line 
and is determined by $\eta_c=1.018828234$, $K=3.27$, and $\beta=0.36$.  We 
obtain black holes with masses down to $M_{\rm BH}\simeq 0.001 M_{\rm ADM}$ 
and the scaling law (8) is seen to hold over two orders of magnitude in 
mass.  Simulations with discretization scales of $3.3\times 10^{-3}$ and 
$1.0\times 10^{-3}$ are included in the plot, using open and filled circles, 
respectively.}
\label{fig3}
\end{figure}

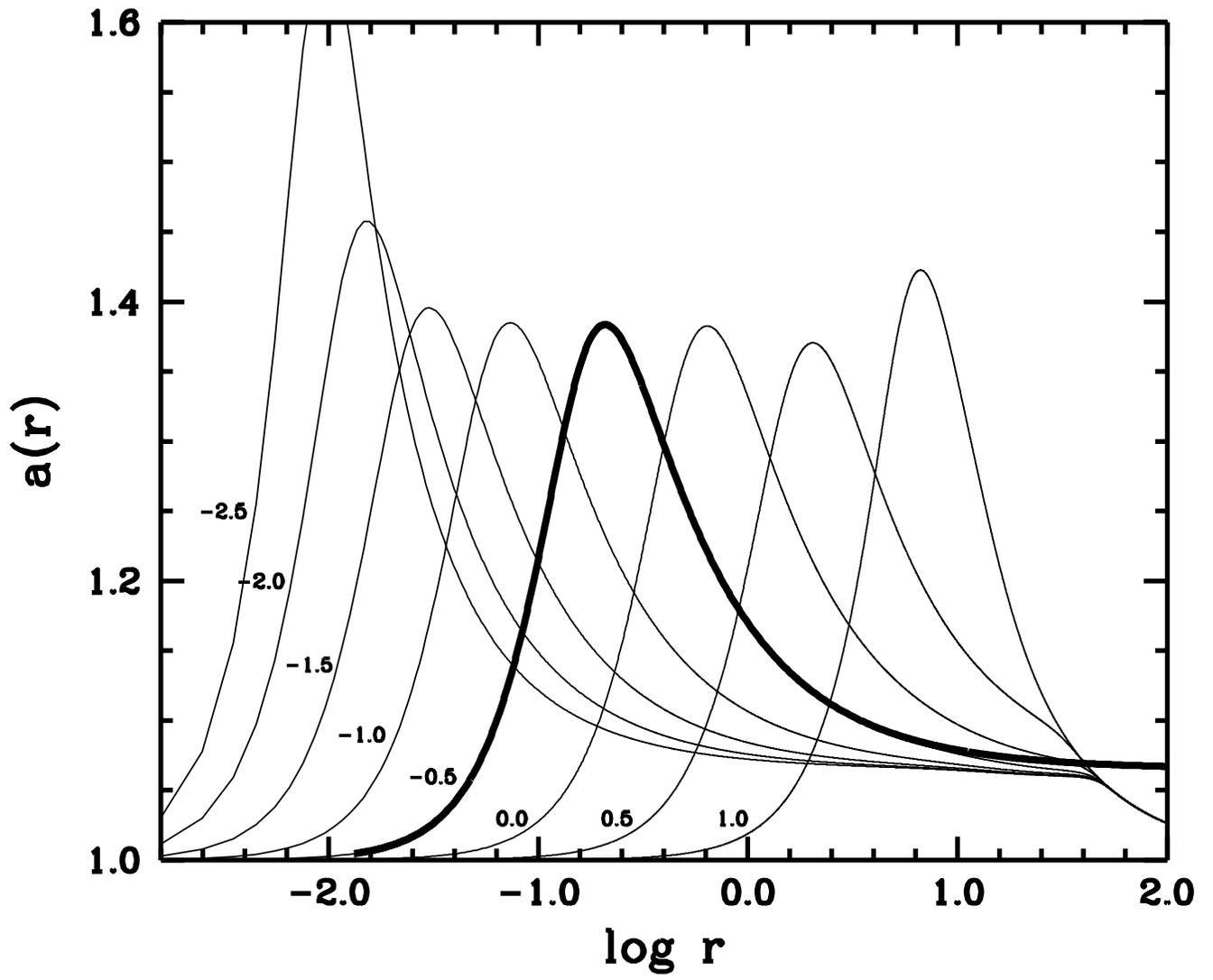
\begin{figure}
\caption{Intermediate asymptotic approach of a near-critical evolution to 
self-similarity.  The metric function $a(r)$ (light curves) at a series of 
times, equally spaced in logarithm of central proper time $\tau=\log(-T)$, 
obtained from a near-critical evolution.  The rightmost curve is near the 
beginning of the simulation and resembles $a(r)$ in the Cauchy data.  As 
collapse proceeds, the peak of $a(r)$ shifts to progressively smaller 
scales.  The metric and fluid approach the self-similar solution depicted 
in Fig.~1.  The self-similar form of $a(\xi)$ (see Fig.~1) is overlaid at 
$\tau=-0.5$, clearly indicating approach to self-similarity.  At later times, 
self-similarity is broken as a black hole begins to form and 
$a\to\infty$ on the horizon.}
\label{fig4}
\end{figure}


\begin{references}
\bibitem{units}We use natural units in which $G=c=1$.
\end{references}
\end{document}